\documentclass[aip,jcp,reprint,superscriptaddress,floatfix]{revtex4-2}
\usepackage{amsmath,amssymb,bm,graphicx,booktabs}
\usepackage[colorlinks=true,citecolor=blue,linkcolor=blue,urlcolor=blue]{hyperref}
\newcommand{\gin}{\gamma_{\mathrm{in}}}
\newcommand{\gout}{\gamma_{\mathrm{out}}}
\newcommand{\Tr}{\operatorname{Tr}}
\newcommand{\ucvaff}{Escuela de F\'isica, Universidad Central de Venezuela, UCV, Caracas 1050, Venezuela}
\newcommand{\RefPPercent}{-0.0340}
\newcommand{\RefSpin}{-0.00337264}
\newcommand{\RefCharge}{9.921297}
\newcommand{\ElasticFactorFive}{1.541}
\newcommand{\ThermalFactorFive}{10.55}
\newcommand{\ThermalPPercent}{-0.443}
\newcommand{\ElasticPPercent}{-0.0652}
\newcommand{\LinearElasticFive}{0.742}
\newcommand{\AbsPPercent}{-0.130}
\newcommand{\ReversePPercent}{0.0340}
\newcommand{\CenteredSpin}{-0.03412272}
\newcommand{\CenteredPPercent}{-0.418}
\newcommand{\OddResidual}{$8.22\times10^{-15}$}
\newcommand{\ProbeResidual}{$1.99\times10^{-11}$}

\begin{document}
\raggedbottom
\title{Spin polarization in a chiral ladder: elastic and inelastic environments}
\author{D. T. Verrilli H.}
\affiliation{\ucvaff}
\author{Jorge Cardenas-Gamboa}
\affiliation{Leibniz Institute for Solid State and Materials Science, 01069 Dresden, Germany}
\affiliation{Max Planck Institute for Chemical Physics of Solids, 01187 Dresden, Germany}
\author{Mayra Peralta}
\affiliation{Max Planck Institute for Chemical Physics of Solids, 01187 Dresden, Germany}
\author{Claudia Felser}
\affiliation{Max Planck Institute for Chemical Physics of Solids, 01187 Dresden, Germany}
\author{Jos\'e H. Garcia}
\affiliation{Catalan Institute of Nanoscience and Nanotechnology (ICN2), CSIC and BIST,
Campus UAB, Bellaterra, 08193 Barcelona, Spain}
\author{Solmar Varela}
\affiliation{\ucvaff}
\author{Nelson Bol\'ivar}
\email{nelson.e.bolivar@ucv.ve}
\affiliation{\ucvaff}
\affiliation{Astrum Drive Technologies, Dallas Pkwy Unit 120 B, Frisco, TX 75034, USA}
\date{17 October 2026}
\begin{abstract}
We study how the transport window and environmental reinjection determine the
spin polarization transmitted through a minimal two-leg chiral ladder. The model
contains time-reversal-invariant helical spin--orbit coupling, spin-independent
interchain hopping, and nonmagnetic contacts with two orbital modes per lead.
Sublattice symmetry makes the coherent spin-transmission kernel exactly odd in
energy. A window centered on the symmetry point therefore gives zero integrated
spin current, whereas a displaced window selects an uncanceled spectral interval.
The resulting polarization reverses between enantiomers and vanishes in the
decoupled-chain, zero-spin--orbit, and equivalent-strand limits of this model.
We compare uniform absorption, energy-resolved elastic probes, and thermalized
voltage probes at the same finite bias, including reinjection whenever probes are
present. Charge conservation alone does not prohibit enhancement of the
integrated spin signal: the elastic-probe response can increase at finite bias
even where its linear-response counterpart decreases. Thermalized probes can
also lift the centered-window zero that elastic probes preserve.
Absorption and reinjection
also change the transmitted charge, so enhancement of spin current and
enhancement of normalized polarization must be distinguished. Static onsite
disorder produces sample-dependent fluctuations; its signed ensemble mean is
reported with uncertainty and is not forced to vanish in a displaced window.
These results identify the energy interval and reservoir prescription that must
be controlled when interpreting dephasing trends in minimal models relevant to
chirality-induced spin selectivity.
\end{abstract}
\maketitle

\section{Introduction}
Spin-selective electron transport through chiral molecular conductors raises a
specific question about the environment: does a change in the measured spin
signal reflect loss of coherence, removal of particles, redistribution in
energy, or a change in the charge used to normalize the signal? The present
work separates these effects in a minimal chiral ladder. Its central result is
a controlled comparison at equal finite bias: a charge-conserving elastic
reservoir prescription can enhance the integrated spin signal even when the
same prescription suppresses the linear-response signal. At the symmetry-centered
window, elastic reinjection preserves an exact zero while thermalized reinjection
can produce a finite signal. The energy window and
the reinjection rule are essential parts of a statement about
environment-assisted spin filtering.

Chirality-induced spin selectivity (CISS) has been observed in several molecular
settings, motivating microscopic and effective descriptions of the relation
between handedness and electron spin.\cite{Ray1999,Kiran2016,Eckvahl2023,Aiello2022,Bloom2024}
Helical tight-binding models isolate symmetry, SOC, and transport channels.
Guo and Sun studied double-helical transport with fictitious
reservoirs,\cite{guo2012} while microscopic orbital projections connect
molecular geometry with effective SOC terms.\cite{Varela2016}
Other approaches include nonunitary scattering, vibrationally induced exchange,
and microscopic spin--charge transport.\cite{Matityahu2016,Fransson2020,Bedoya2026}
Interference and contact placement also affect spin selection in helical
loops.\cite{ChenFu2023}

Time-reversal symmetry (TRS) imposes Kramers degeneracy of transmission
eigenvalues in coherent two-terminal transport.\cite{Bardarson2008}
It does not prohibit all spin polarization when several orbital channels are
available: Utsumi, Entin-Wohlman, and Aharony demonstrated coherent multichannel
spin filtering explicitly.\cite{Utsumi2020}
Symmetry constraints concern the entire junction, including its contacts.
In particular, TRS combined with particle--hole symmetry can make a spectral
spin conductance odd in energy.\cite{Dednam2023} We use these established
restrictions rather than identifying coherent multichannel filtering or
spectral oddness as new effects.

Our comparison keeps the device Hamiltonian, contact channels, and transport
window fixed while changing the reservoir prescription. Uniform absorption
is a loss-only reference. Elastic probes enforce zero charge current at each
energy; voltage probes enforce zero integrated charge current with a Fermi
distribution in each probe.\cite{buttiker1986role,DAmato1990}
Both reintroduce particles, but only the latter permits energy redistribution.
Huisman and Thijssen already distinguished these probe types and loss-only
models when studying nonlinear magnetoresistance in helicene
junctions.\cite{Huisman2021} Here the observable is the outgoing spin flux
with nonmagnetic leads, rather than a magnetic-detector conductance difference.
Neither imposes zero spin current in a probe. This distinction is needed to
interpret a calculation labeled as dephasing.

The ladder is a qualitative model, not a parametrization of DNA or of a
particular contact chemistry. We calculate spectral and integrated polarization,
compare complete enantiomeric junctions, and display the length dependence of
charge and polarization separately. These observables establish its connection
with chiral spin filtering while specifying which aspects of molecular CISS
remain outside its scope.

\section{Model and transport observables}
\subsection{Time-reversal-invariant chiral ladder}
The device has two legs, $j=A,B$, each containing $N$ sites and two spin states
per site. Its single-particle Hilbert space has dimension $4N$. With
$c_{j,n}=(c_{j,n,\uparrow},c_{j,n,\downarrow})^T$, the Hamiltonian is
\begin{align}
H={}&\sum_{j,n}\epsilon_{j,n}c_{j,n}^{\dagger}c_{j,n}
 +\gout\sum_n(c_{A,n}^{\dagger}c_{B,n}+\mathrm{h.c.})\nonumber\\
 &+\sum_{j,n=1}^{N-1}(c_{j,n}^{\dagger}B_{j,n}c_{j,n+1}
 +\mathrm{h.c.}),\label{eq:H}\\
B_{j,n}={}&\gin\mathbf{1}_2-i\lambda_{\mathrm{soc}}
 [\cos\phi_{j,n}\,\sigma_x+\sin\phi_{j,n}\,\sigma_y].\label{eq:bond}
\end{align}
Onsite energies vanish in the clean model. Interchain hopping is real and
spin independent. The helical angles are
\begin{equation}
\begin{split}
\phi_{A,n}&=\chi n\Delta\phi,\quad
\phi_{B,n}=\chi(n\Delta\phi+\beta),\\
\Delta\phi&=\frac{2\pi}{M-1},\qquad \chi=\pm1.
\end{split}\label{eq:angles}
\end{equation}
Here $\chi$ labels handedness and $\beta$ the relative strand azimuth.
The SOC axis completes a $2\pi$ rotation in $M-1$ longitudinal steps.
Both strands co-rotate.

\begin{figure*}[tbp]
\centering
\includegraphics[width=6.69in]{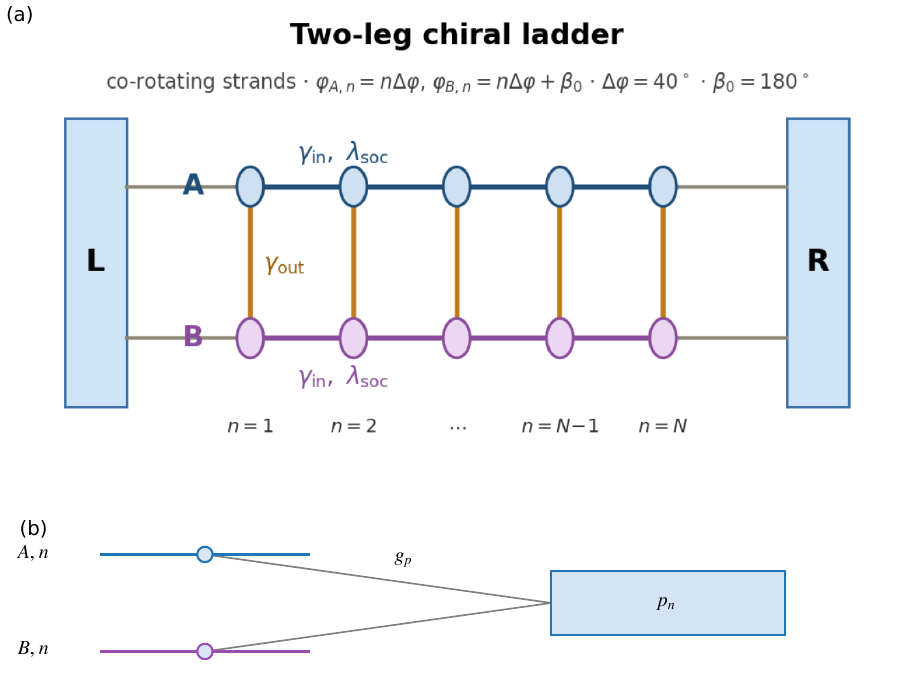}
\caption{(a) The two-leg chiral ladder. The diagram uses $\beta_0=\beta$
and shows the reference handedness, $\Delta\phi=40^\circ$ and $\beta=180^\circ$.
Both legs contact independent orbital modes of each nonmagnetic reservoir.
(b) Environmental coupling for one rung. A single probe $p_n$ couples to
all four states $(A\uparrow,A\downarrow,B\uparrow,B\downarrow)$ with halfwidth
$g_p$ and a common spin-independent distribution.}\label{fig:model}
\end{figure*}

Equation~\eqref{eq:bond} is an effective radial SOC ansatz. It retains a
rotating spin axis and the oriented imaginary hopping required by TRS;
it is not asserted to be a term-by-term projection of the multiorbital
Hamiltonian of Ref.~\onlinecite{Varela2016}. It omits additional orbital,
longitudinal-spin, and geometry-dependent terms of microscopic descriptions.
Its symmetry is explicit: for $\Theta=\mathbf{1}_{2N}\otimes i\sigma_yK$,
$\sigma_yB_{j,n}^{*}\sigma_y=B_{j,n}$ and $\Theta H\Theta^{-1}=H$.
Replacing $-i\lambda_{\mathrm{soc}}$ by a real spin-dependent hopping breaks
this relation.
We separate $H=H_0+H_{\mathrm{so}}$: $H_0$ contains the onsite energies and
all spin-independent hoppings, including $\gout$, whereas $H_{\mathrm{so}}$
contains the terms proportional to $\lambda_{\mathrm{soc}}$.
Restrictions on spin-dependent interstrand hopping must therefore be
distinguished from restrictions on the scalar rung term in $H_0$.
For equivalent legs at $\beta=\pi$, TRS and a half-turn about the helix axis
permit $\gout\mathbf{1}_2$ in the effective basis used here.
Appendix~\ref{app:rung} derives this statement and identifies the allowed
spin-dependent rung terms, which are set to zero in the present model.
Line-group-based microscopic descriptions address an explicit orbital
representation.\cite{Mendieta2026} Our symmetry check does not establish
a microscopic orbital projection or equivalence to those Hamiltonians.

Co-rotating strands also appear in Ref.~\onlinecite{guo2012}, whose SOC
matrices involve azimuths $\phi$ and $\phi+\pi$. The present model differs
in SOC components, onsite energies, hoppings, and independent orbital modes
of the contacts. A difference between coherent limits cannot be attributed
to strand handedness alone. Oppositely rotating strands are used below
only as a symmetry control.

\subsection{Contact widths, Green functions, and polarization}
The contacts act on the first and last rungs with projectors $P_L,P_R$.
Their wide-band self-energies and level-width matrices are
\begin{equation}
\Sigma_q^r=-ig_qP_q,\quad
\Gamma_q=i(\Sigma_q^r-\Sigma_q^a)=2g_qP_q,\quad q=L,R.
\label{eq:width}
\end{equation}
Thus $g_q$ is a halfwidth. The equation-of-motion hierarchy for the quadratic
Hamiltonian closes exactly, without an interaction decoupling, and gives
\begin{equation}
G^r(E)=[E+i0^+-H-\Sigma_L^r-\Sigma_R^r-\Sigma_d^r]^{-1}.
\label{eq:G}
\end{equation}
This is the usual retarded resolvent.\cite{Datta1995,Haug2008}
For unpolarized injection from $L$, the charge and outgoing $z$-spin kernels are
\begin{equation}
T^\alpha(E)=\Tr[\sigma_R^\alpha\Gamma_RG^r\Gamma_LG^a],
\quad \alpha=0,z.\label{eq:caroli}
\end{equation}
Here $\sigma^0=\mathbf{1}$ and $\sigma^z$ acts on the outgoing spin. Equivalently,
\begin{equation}
T^\alpha=4g_Lg_R\sum_{a\in R,b\in L}s_a^\alpha|G^r_{ab}|^2,
\quad s_a^z=\begin{cases}+1,&a\uparrow,\\-1,&a\downarrow.
\end{cases}\label{eq:channels}
\end{equation}
The prefactor follows from the full widths in Eq.~\eqref{eq:width}.
Four incident spin-orbital channels imply $0\leq T^0\leq4$ and
$|T^z|\leq T^0$. The upper bound also holds for the subunitary loss-only
transmission, but does not alone establish conservation.

Define bias energy $\Delta\mu=eV$ and $\mu_{L,R}=E_F\pm\Delta\mu/2$.
For a fixed Hamiltonian at zero temperature,
\begin{equation}
I_\alpha=\frac{e}{h}\int_{\mu_R}^{\mu_L}T^\alpha(E)\,dE,
\quad P(E)=\frac{T^z(E)}{T^0(E)},\quad
P_{\mathrm{int}}=\frac{I_z}{I_0}.\label{eq:current}
\end{equation}
The charge denominator must be nonzero. $I_z$ is the difference of
spin-resolved charge currents; the angular-momentum flux is $\hbar I_z/(2e)$.
Spin is generally not conserved inside the SOC region.\cite{ShiNiuPRL2006}
For the absorber, Eq.~\eqref{eq:current} defines a cross-terminal transmitted
signal of the loss-only reference. Probes require the complete reservoir
expression of Sec.~\ref{sec:environment}.

Reference parameters are $\gin=\gout=g_L=g_R=1$ meV,
$\lambda_{\mathrm{soc}}=0.1$ meV, $N=M=10$, and $\beta=\pi$.
The window has $E_F=0.5$ meV and $\Delta\mu=4$ meV, giving
$[\mu_R,\mu_L]=[-1.5,2.5]$ meV. This is a model operating point, not a fitted
molecular alignment. The internal potential is held fixed under bias.
Figure currents are in units $I_*=e\gin/h$.

\section{Symmetry and the transport window}
\subsection{An exact spectral cancellation}
For zero onsite energies, define the sublattice operator, with $j=0,1$ on $A,B$,
\begin{equation}
S|n,j,s\rangle=(-1)^{n+j}|n,j,s\rangle,\qquad SHS=-H.
\label{eq:S}
\end{equation}
Every longitudinal bond and rung connects opposite sublattices. Alternating
only in $n$ would leave the rung term unchanged. The purely imaginary diagonal
self-energies commute with $S$, giving
\begin{equation}
G^r(-E)=-SG^a(E)S.\label{eq:chiralG}
\end{equation}
Combining this identity with spin-$1/2$ reciprocity yields
\begin{equation}
T^0(-E)=T^0(E),\qquad T^z(-E)=-T^z(E).\label{eq:odd}
\end{equation}
Appendix~\ref{app:symmetry} supplies the intermediate steps and assumptions.
The same kernel identity holds with a scalar diagonal absorber, but does
not specify the occupations injected by additional reservoirs.

\begin{figure*}[tbp]
\centering
\includegraphics[width=6.69in]{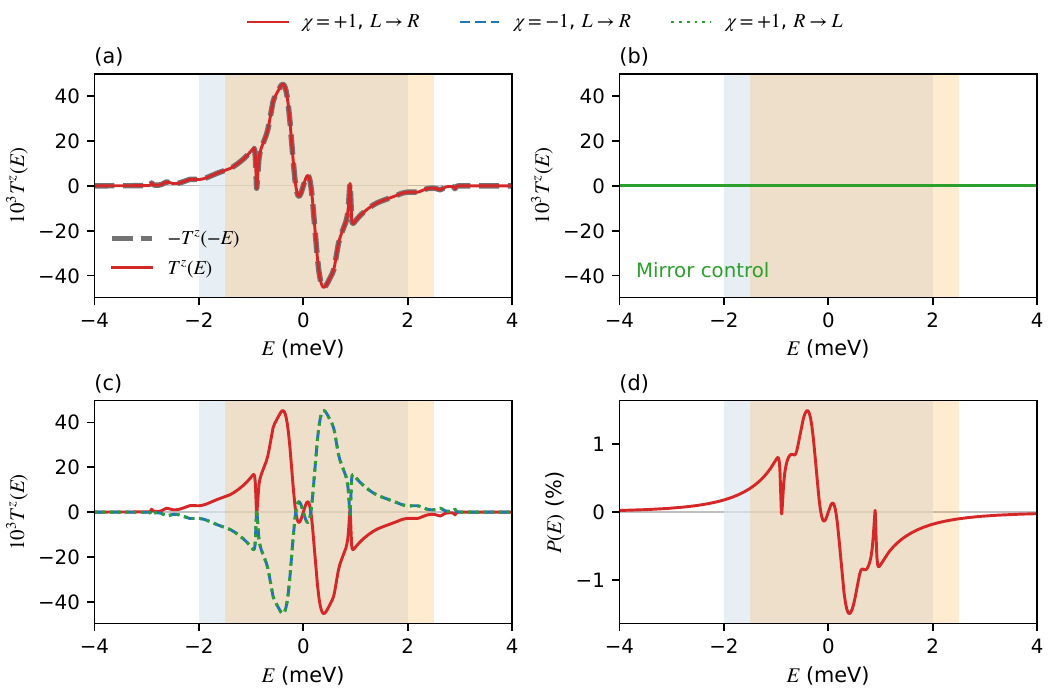}
\caption{Spectral spin transport. (a) The reference kernel $T^z(E)$ (red)
overlaps $-T^z(-E)$ (gray dashed), verifying its odd parity on the plotted grid.
(b) The mirror-control kernel vanishes. (c) Kernels for both enantiomers
and reversed propagation of the reference helix. (d) Normalized spectral
polarization. Blue and orange shading mark the centered $[-2,2]$ meV and
displaced $[-1.5,2.5]$ meV windows, respectively; their overlap mixes the
colors. The interval $[-1.5,1.5]$ meV cancels from the displaced-window spin
integral, leaving $[1.5,2.5]$ meV.}
\label{fig:spectral}
\end{figure*}

Figure~\ref{fig:spectral} shows the odd spin kernel, its mirror-control zero,
and the normalized polarization. Thus $I_z(E_F=0)=0$ exactly for any bias energy in
the fixed-Hamiltonian coherent model. The zero follows from spectral symmetry
and the window, rather than from Kramers degeneracy alone. The computed
maximum oddness residual is \OddResidual; it verifies Eq.~\eqref{eq:odd}
on the stated grid and is not its proof.
For any odd kernel,
\begin{equation}
\int_{E_F-\Delta\mu/2}^{E_F+\Delta\mu/2}T^z(E)\,dE
=\int_{\Delta\mu/2-E_F}^{\Delta\mu/2+E_F}T^z(E)\,dE.
\label{eq:slice}
\end{equation}
Hence $I_z(-E_F)=-I_z(E_F)$. Equation~\eqref{eq:slice} also identifies the
surviving spectral interval: $[1.5,2.5]$ meV for the reference window.
Linear response instead probes $T^z(E_F)$ at $0.5$ meV. Their environmental
trends need not have the same sign.

\subsection{Geometry, enantiomers, and necessary ingredients}\label{sec:geometry}
The mirror control has $\phi_{B,n}=-\phi_{A,n}+\beta_m$. Let
$F_{\beta_m}=\cos(\beta_m/2)\sigma_x+\sin(\beta_m/2)\sigma_y$ and let
$\mathcal P$ exchange the legs. Then
$F_{\beta_m}B(\phi)F_{\beta_m}=B(\beta_m-\phi)$.
For equivalent legs and contacts, $U=\mathcal PF_{\beta_m}$ commutes with
$H$ and the contact widths, while $U\sigma_zU^\dagger=-\sigma_z$.
The spin trace equals its negative, so $T^z(E)=0$ at every energy.
This is a property of that symmetric junction, not all double-helical conductors.

Table~\ref{tab:controls} contrasts this zero with the co-rotating reference.
Removing SOC, rung hopping, or relative offset also removes the spin kernel
in the reference family while charge transmission remains finite. These
conditions concern the stated Hamiltonian and contacts, not every CISS mechanism.
\begin{table*}[tbp]
\caption{Controls at $N=10$, $E_F=0.5$ meV, $\Delta\mu=4$ meV.
Currents are in units $I_*$. The mirror case has $\beta_m=\pi$.
The supplementary data give sampled spectral residuals for the numerical zeros.}
\label{tab:controls}
\centering\begin{tabular}{lrr}
\toprule
Model & $I_0/I_*$ & $I_z/I_*$\\
\midrule
Co-rotating & 9.92130 & -0.00337264\\
$\lambda_{\mathrm{soc}}=0$ & 9.92380 & $\lvert I_z/I_*\rvert<10^{-14}$\\
$\gamma_{\mathrm{out}}=0$ & 11.63674 & $\lvert I_z/I_*\rvert<10^{-14}$\\
$\beta=0$ & 9.96036 & $\lvert I_z/I_*\rvert<10^{-14}$\\
Mirror control & 9.92467 & $\lvert I_z/I_*\rvert<10^{-14}$\\
\bottomrule
\end{tabular}

\end{table*}

\begin{figure}[tbp]
\centering
\includegraphics[width=3.37in]{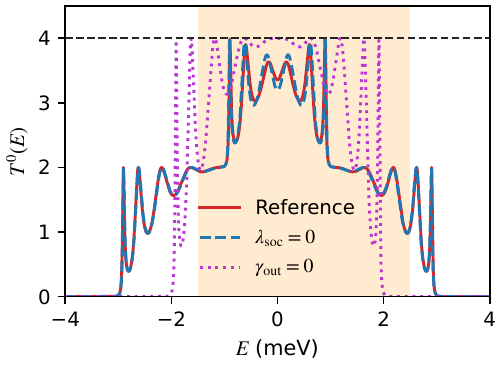}
\caption{Charge-transmission controls at $N=M=10$. The reference ladder
(red), zero-SOC case (blue dashed), and decoupled legs (magenta dotted)
all transmit charge. The black dashed line is the four-channel bound
$T^0=4$; orange shading marks the displaced window $[-1.5,2.5]$ meV.}
\label{fig:charge}
\end{figure}
Figure~\ref{fig:charge} displays the corresponding charge kernels, which are
even in energy. Charge transmission survives when SOC or the rung hopping
is removed, although the spin kernel vanishes in these controls. Thus a
charge-transmission curve alone does not establish the ingredients needed
for a spin-selective response.

Reflecting the complete co-rotating junction changes $\chi$ and hence every
helical angle. A global $\sigma_x$ conjugation maps the two Hamiltonians
into one another and reverses $\sigma_z$. For the same contacts and fixed
laboratory spin axis, $T^0$ is unchanged and $T^z$ changes sign. The reference
$P_{\mathrm{int}}=\RefPPercent\%$ has the opposite sign to its enantiomer.
This compares complete junctions rather than opposite strands inside one device.

\begin{figure*}[tbp]
\centering
\includegraphics[width=6.69in]{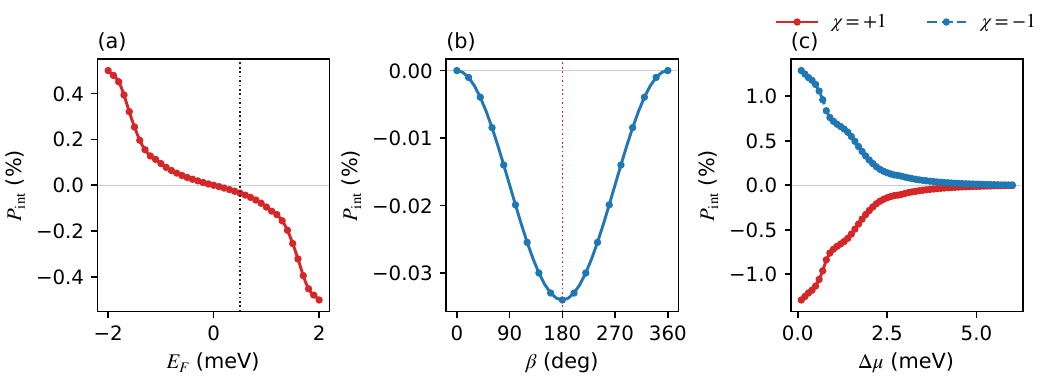}
\caption{Integrated polarization versus (a) window center at
$\Delta\mu=4$ meV, (b) relative strand angle at $E_F=0.5$ meV, and
(c) bias energy for both enantiomers at $E_F=0.5$ meV.
Dotted vertical lines in (a,b) mark the reference $E_F=0.5$ meV and
$\beta=180^\circ$. All remaining parameters are those of the reference device.}
\label{fig:controls}
\end{figure*}
Figure~\ref{fig:controls} separates alignment, geometry, and bias. The plotted
normalized polarization grows in magnitude over the sampled positive offsets;
spectral oddness alone does not imply a global monotonicity law.
The angle response vanishes at
$\beta=0,2\pi$ and has its largest sampled magnitude near $\pi$ for this
parameter set. This is a maximum of magnitude, not of the negative signed current.

\section{Absorption and conserving reinjection}\label{sec:environment}
\subsection{Three reservoir prescriptions at equal bias}
The loss-only reference has $\Sigma_d^r=-i\eta_d\mathbf{1}_{4N}$ and no
reinjection. It broadens the retarded resolvent and removes carriers.
For a conserving construction, one reservoir couples to each rung with
$\Sigma_p^r=-ig_pP_p$ and $\Gamma_p=2g_pP_p$. All four states share its
spin-independent distribution. This permits spin relaxation; zero charge
current does not imply zero spin current into a probe.

For the physical and probe terminals the unitary scattering matrix is
\begin{equation}
\mathcal S(E)=\mathbf{1}-iW^\dagger G^r(E)W,\qquad
WW^\dagger=\sum_q\Gamma_q.\label{eq:scattering}
\end{equation}
Let $\mathcal T_{pq}=\Tr(\mathcal S_{pq}\mathcal S_{pq}^\dagger)$ and
$\mathcal Z_{pq}=\Tr(\sigma_z\mathcal S_{pq}\mathcal S_{pq}^\dagger)$,
including reflection for $p=q$, and define
$\mathcal M_{pq}=4\delta_{pq}-\mathcal T_{pq}$. The charge current leaving
terminal $p$ and spin signal leaving the device into $R$ are
\begin{align}
I_p^{\mathrm{out}}&=\frac{e}{h}\int dE\sum_q\mathcal M_{pq}(E)f_q(E),
\label{eq:LBcharge}\\
I_{z,R}&=\frac{e}{h}\int dE\sum_q\mathcal Z_{Rq}(E)f_q(E).
\label{eq:LBspin}
\end{align}
Incoming distributions are unpolarized. Unitarity cancels the equilibrium
spin contribution when all $f_q$ are equal. At a conserving solution,
$I_0=I_L^{\mathrm{out}}=-I_R^{\mathrm{out}}$ and
$P_{\mathrm{int}}=I_{z,R}/I_0$.

An elastic probe imposes $\sum_q\mathcal M_{pq}(E)f_q(E)=0$ at each energy.
It conserves charge and energy flux. A thermalized voltage probe instead
has $f_p(E)=f(E;\mu_p,T)$, and $\mu_p$ solves
$I_p^{\mathrm{out}}=0$ after integration. It conserves charge but exchanges
energy with its environment. All results below use $T=0$ and the same
$[-1.5,2.5]$ meV physical-lead window. Equal numerical $g_p=\eta_d$
gives the same added retarded damping; occupation and reinjection distinguish
the prescriptions.

\subsection{Finite-bias and linear-response trends}
\begin{figure*}[tbp]
\centering
\includegraphics[width=6.69in]{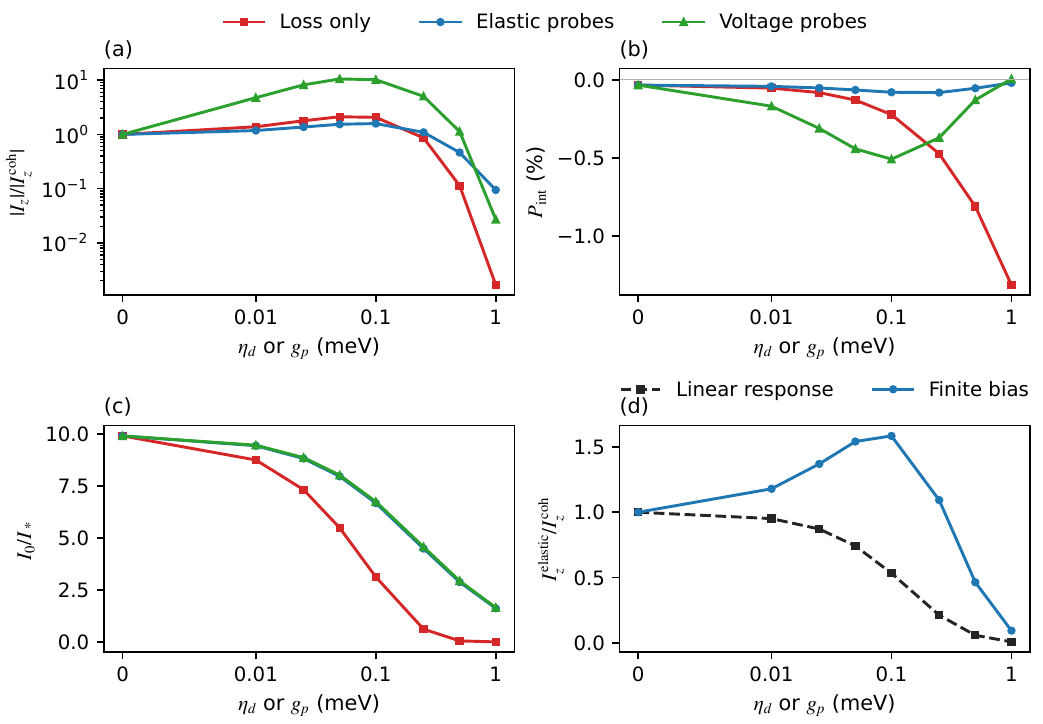}
\caption{Environmental comparison at equal bias: (a) spin-current magnitude
relative to its coherent value, (b) normalized polarization, (c) charge,
and (d) elastic-probe spin-current ratio in linear response and at finite
bias. Panel (a) uses a logarithmic vertical scale. Multiplying a
linear-response kernel by $\Delta\mu$ is an approximation,
not a finite-window integral. Only the loss-only curve omits reinjection.}
\label{fig:environment}
\end{figure*}

Figure~\ref{fig:environment} resolves the distinction. The coherent reference
has $I_z/I_*=\RefSpin$ and $I_0/I_*=\RefCharge$.
At $g_p=0.05$ meV, elastic probes give a finite-bias spin-current magnitude
\ElasticFactorFive\ times the coherent value. At the same operating point,
the linear-response elastic-probe ratio is \LinearElasticFive. This is a conserving
example of enhancement after integration, without implying enhancement of
every spectral component or every bias regime.

Loss-only and thermalized-probe results use the same coupling convention.
At $g_p=0.05$ meV, thermalized probes give a spin-current magnitude
\ThermalFactorFive\ times the coherent value and
$P_{\mathrm{int}}=\ThermalPPercent\%$, compared with
$\ElasticPPercent\%$ for elastic probes. These are calculated values at the
specified operating point, not maxima over continuous parameters.
Their trends follow from different occupations and spectral weights; they
cannot be ordered solely by the label dephasing. In particular, monotonic
linear-response suppression does not establish that a finite-bias increase
is exclusive to an absorber. Adaptive quadrature of the thermalized solutions
gives a maximum probe-current residual of \ProbeResidual\ in units $I_*$.
Numerical methods, refinement, and disorder statistics are specified in the
supplementary material. Tables S3 and S4 there report the sampled
elastic-probe charge residuals and integrated voltage-probe residuals,
respectively, together with an independent current evaluation.

Charge also changes appreciably. The maxima of $|I_z|$ and
$|P_{\mathrm{int}}|$ need not coincide. At $\eta_d=0.05$ meV, the loss-only
reference gives $P_{\mathrm{int}}=\AbsPPercent\%$, compared with
$\RefPPercent\%$ coherently. The polarization ratio differs from the
spin-current enhancement ratio. Attenuation of long interfering paths is
an intuitive interpretation of spectral reshaping; no trajectory-resolved
decomposition is assumed in the calculation.

\subsection{When energy redistribution lifts the centered-window zero}
The symmetry-centered window supplies a separate test of the reinjection rule.
For the onsite-free Hamiltonian and the stated spin-independent contacts,
the spectral identities extend to every block of the enlarged scattering matrix:
\begin{equation}
\mathcal M_{pq}(-E)=\mathcal M_{pq}(E),\quad
\mathcal Z_{pq}(-E)=-\mathcal Z_{pq}(E).
\label{eq:probeparity}
\end{equation}
Inside a centered zero-temperature window, elastic-probe occupations solve
\begin{equation}
\bm f_P(E)=-\mathcal M_{PP}^{-1}(E)\mathcal M_{PL}(E),
\quad f_L=1,\quad f_R=0.
\label{eq:elasticdistribution}
\end{equation}
For connected probes this solution is even in energy. Its product with the
odd spin row therefore integrates to zero exactly. An arbitrary centered
physical-lead window does not impose this evenness on the distinct Fermi steps
$f_p(E)=\theta(\mu_p-E)$ of thermalized probes.

Table~\ref{tab:centered} shows the resulting distinction. At $E_F=0$ and
$g_p=0.05$ meV, the computed thermalized signal is
$I_{z,R}/I_*=\CenteredSpin$, with $P_{\mathrm{int}}=\CenteredPPercent\%$.
The enantiomer has the opposite sign and equal charge. The comparison
isolates energy redistribution as the ingredient that lifts this particular
centered-window cancellation; it does not make inelasticity necessary for
every chiral spin-filtering response. In particular, the displaced-window
coherent response remains finite. Neither probe prescription is a microscopic
electron--phonon self-energy.
\begin{table}[!htbp]
\caption{Centered window $[-2,2]$ meV, $N=M=10$ and reference couplings.
The environmental halfwidth is $0.05$ meV. Zero-current entries are numerical
checks of exact cancellations; the last row is a calculated finite response.}
\label{tab:centered}
\centering\begin{tabular}{lrrr}
\toprule
Prescription & $I_0/I_*$ & $I_z/I_*$ & $P_{\rm int}$ (\%)\\
\midrule
Coherent & 10.06910 & $<10^{-14}$ & $<10^{-12}$\\
Loss only & 5.62157 & $<10^{-14}$ & $<10^{-12}$\\
Elastic probes & 8.09897 & $<10^{-14}$ & $<10^{-12}$\\
Voltage probes & 8.15910 & -0.034123 & -0.418\\
\bottomrule
\end{tabular}

\end{table}

\section{Length and static disorder}
\begin{figure*}[tbp]
\centering
\includegraphics[width=6.69in]{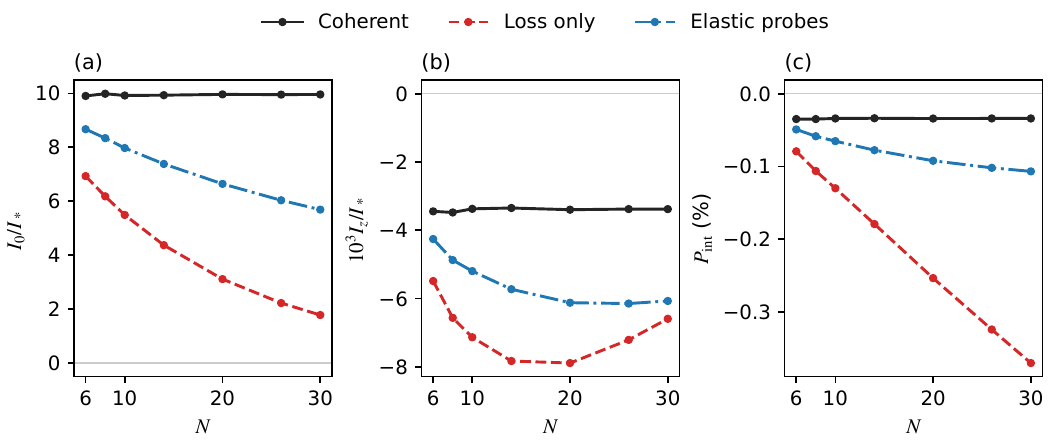}
\caption{Length dependence at fixed $M=10$, $E_F=0.5$ meV and
$\Delta\mu=4$ meV. Charge, spin signal, and polarization are calculated
separately for the coherent ladder and indicated environments. An
environmental halfwidth of $0.05$ meV is used for both reservoir prescriptions.
A smaller
spin current does not alone imply smaller polarization. Lines connect
sampled lengths.}\label{fig:length}
\end{figure*}
Figure~\ref{fig:length} compares charge and polarization directly instead
of inferring the latter from $I_z$. The response concerns this multichannel
ladder and its alignment. It is not an asserted monotonic length law for
molecular CISS, nor does declining charge imply declining filtering efficiency.

For disorder, the $\epsilon_{j,n}$ are independent uniform samples from
$[-W/2,W/2]$. The coherent device is evaluated for every realization before
averaging. At $\beta=\pi$, strand exchange combined with $\sigma_z$ maps
a pattern onto its strand-interchanged partner while preserving the measured
$z$ spin. That exchange does not force a zero signed ensemble mean.

\begin{figure*}[tbp]
\centering
\includegraphics[width=6.69in]{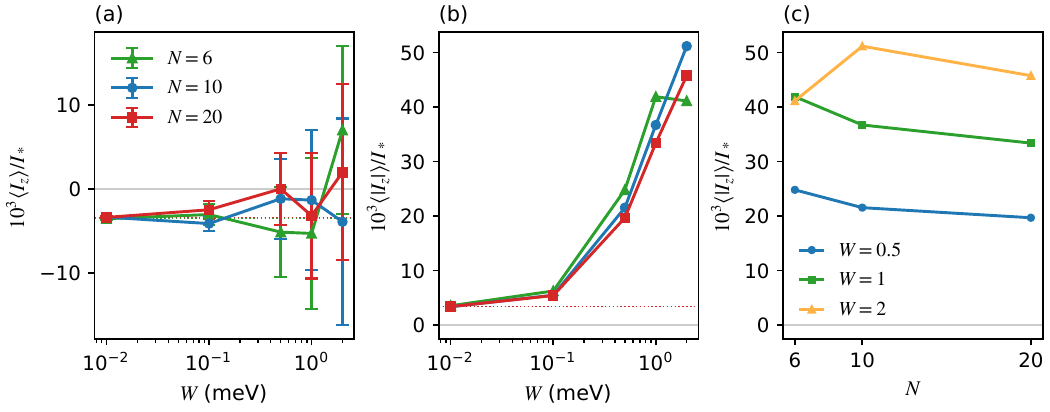}
\caption{Static onsite disorder at the reference window. (a) Signed mean
spin current with 95\% Student-$t$ intervals from 128 independent realizations
at each point. (b) Mean absolute current, a distinct statistic that can remain
large when the signed mean is poorly resolved. Clean references are indicated.
(c) The same mean absolute currents plotted versus length for
$W=0.5,1,2$ meV.
Onsite arrays, individual currents, bootstrap intervals, and charge-normalized
statistics accompany the supplementary data.}\label{fig:disorder}
\end{figure*}
Figure~\ref{fig:disorder} separates signed and absolute statistics. Panel (c)
also displays the length dependence of the absolute means at three disorder
strengths, without replacing the signed statistic in panel (a). Weak
disorder connects continuously to the nonzero clean response. At larger
disorder, sample variations can exceed that clean signal. An interval
containing zero reflects limited precision for a sampled mean; it does not
prove a zero population mean. A symmetric onsite distribution does impose
an ensemble zero for a centered window under the assumptions in
Appendix~\ref{app:symmetry}; it does not force a zero in the displaced window.

\section{Connection with CISS and limitations}
The enantiomer comparison establishes chirality-dependent spin filtering
in this model. Its reference polarization is small. Comparison with molecular
data requires matching the regime, injection modes, spin analyzer, and contacts.
A transmitted-flux polarization is not automatically a magnetoconductance
asymmetry measured with a ferromagnetic detector.\cite{Jacquod2012,Huisman2021}

Interchanging source and detector at fixed laboratory $z$ axis changes the
reference polarization to $\ReversePPercent\%$. Thus this model does not
account for measurements whose detector signal retains its sign under current
reversal. Normalizing the current alone cannot supply the missing detection
physics or other mechanisms.

Coherent multichannel filtering and spectral parity are established
results.\cite{Utsumi2020,Dednam2023} The contribution of the present comparison
is to hold the Hamiltonian, widths, and finite-bias window fixed while
separating loss, energy-resolved reinjection, and thermalized reinjection.
It identifies the surviving spectral interval and demonstrates that a
linear-response environmental trend can change after integration.
For a centered window it also separates the exact cancellation with elastic
probes from the finite response permitted by thermalized occupations.
The calculation does not resolve the general microscopic origin of CISS.

\section{Conclusions}
The spin signal of a chiral ladder depends on which spectral contributions
the reservoirs populate. With the stated contacts, an exactly odd coherent
spin kernel integrates to zero in a centered window. A displaced window
retains a specific uncanceled interval and gives polarization that reverses
between enantiomers.

At fixed finite bias, charge-conserving elastic reinjection can enhance the
integrated signal where its linear-response counterpart decreases.
Absorption, elastic reinjection, and thermalized reinjection must therefore
be compared at a common bias before assigning a trend to dephasing. The
associated charge is needed to establish polarization. At the symmetry-centered
window, elastic probes retain the exact zero, whereas thermalized probes
can populate the spin channels unequally after integration. That distinction
concerns reservoir occupations, not a loss of TRS in the device Hamiltonian.
Likewise, a larger
typical disorder response does not establish a systematic signed signal.
These distinctions provide a reproducible baseline for assessing microscopic
CISS mechanisms and realistic molecular contacts.

\section*{Supplementary Material}
The supplementary material provides reservoir-current derivations, probe
potentials, tabulated observables, conservation residuals, numerical
refinement, disorder statistics, and additional current plots. The
accompanying numerical archive contains the transport implementation,
individual disorder realizations, retained currents, and figure inputs.
\section*{Acknowledgments}
This research received no specific grant from any funding agency in the
public, commercial, or not-for-profit sectors.
\section*{Author Declarations}
\subsection*{Conflict of Interest}
The authors have no conflicts to disclose.
\subsection*{Author Contributions}
All authors: Formal analysis (equal); Investigation (equal);
Writing (original draft, equal); Writing (review and editing, equal).
D. T. Verrilli H., Jorge Cardenas-Gamboa, Jos\'e H. Garcia, and
Nelson Bol\'ivar: Software.

\section*{Data Availability}
The data and code that support the findings of this study are available
in the supplementary material, including its numerical archive, and
correspond to the Hamiltonian, contacts, and windows specified here.

\appendix
\section{Spectral identities and their domain}\label{app:symmetry}
For $Y=\mathbf{1}_{2N}\otimes\sigma_y$, the onsite-free ladder and its
diagonal imaginary self-energies obey
\begin{align}
G^r(-E)&=-SG^a(E)S,\\
G^r(E)&=Y[G^r(E)]^TY,\\
G^a(E)&=Y[G^r(E)]^*Y.\label{eq:reciprocity}
\end{align}
The contact widths are real, spin independent, and commute with $S$.
Substitution in Eq.~\eqref{eq:caroli}, conjugation of the real transmission
trace, and $Y\sigma_zY=-\sigma_z$ give Eq.~\eqref{eq:odd}.
These steps also hold for a scalar diagonal absorber. Arbitrary onsite
energies or incompatible contact matrices need not preserve the identity.
For the enlarged scattering problem, assign to each terminal the restriction
$D_q$ of $S$ to its channels. Since the couplings preserve sublattice and spin,
Eq.~\eqref{eq:chiralG} gives
$\mathcal S_{pq}(-E)=D_p\mathcal S_{qp}^{\dagger}(E)D_q$.
Spin-$1/2$ reciprocity replaces $\mathcal S_{qp}^{\dagger}$ by
$Y_p\mathcal S_{pq}^{*}Y_q$.
The $D_q$ commute with the spin operators, while
$Y_q\sigma_zY_q=-\sigma_z$. Charge and spin traces then give
Eq.~\eqref{eq:probeparity}, including reflection blocks.

For onsite pattern $\epsilon$, sublattice conjugation relates two devices,
$SH(\epsilon)S=-H(-\epsilon)$. Thus
$T^z(-E;-\epsilon)=-T^z(E;\epsilon)$. A distribution symmetric under
$\epsilon\to-\epsilon$ gives
$\langle I_z(-E_F)\rangle=-\langle I_z(E_F)\rangle$ and a zero ensemble
mean at $E_F=0$, but no forced zero at fixed nonzero $E_F$.
Individual disorder realizations need not have an odd kernel.
The supplementary material compares explicit sign-inverted partners and
distinguishes these constructed controls from the independent samples
used in the confidence intervals.

\section{Scalar and spin-dependent interstrand hopping}\label{app:rung}
Consider equivalent legs at relative azimuth $\beta=\pi$ in the effective
rung basis $(A\uparrow,A\downarrow,B\uparrow,B\downarrow)$.
Let $\tau_\alpha$ act on the leg index. A half-turn about $z$ exchanges the
legs and rotates the spin, with representation
\begin{equation}
U_2=-i\tau_x\otimes\sigma_z.
\end{equation}
In the real effective orbital basis, write the Hermitian rung Hamiltonian as
\begin{equation}
H_\perp=\begin{pmatrix}0&C_\perp\\C_\perp^\dagger&0\end{pmatrix}.
\end{equation}
TRS and $U_2H_\perp U_2^\dagger=H_\perp$ impose, respectively,
\begin{equation}
C_\perp=\sigma_yC_\perp^*\sigma_y,\qquad
C_\perp=\sigma_zC_\perp^\dagger\sigma_z.
\end{equation}
Expanding a general complex $2\times2$ matrix in Pauli matrices, the first
condition gives $C_\perp=a\mathbf{1}_2+i\bm b\cdot\bm\sigma$, with real
$a$ and $\bm b$. The second removes $b_z$ but not $a$, leaving
\begin{equation}
C_\perp=a\mathbf{1}_2+i b_x\sigma_x+i b_y\sigma_y,
\quad a,b_x,b_y\in\mathbb R.\label{eq:allowedrung}
\end{equation}
Thus the scalar choice $a=\gout$, $b_x=b_y=0$ is symmetry allowed.
The transverse spin-dependent terms are also allowed, but are omitted in
our restricted effective model. Their omission is not a requirement of
the two symmetries considered here.

For the longitudinal bonds at $\beta=\pi$,
$B_{B,n}=\sigma_z B_{A,n}\sigma_z$. Hence the clean Hamiltonian and the
leg-equivalent contact and probe widths share the half-turn symmetry.
This symmetry preserves the measured spin,
$U_2(\mathbf{1}_2\otimes\sigma_z)U_2^\dagger
=\mathbf{1}_2\otimes\sigma_z$, so it does not force a zero spin kernel.
It differs from the mirror control of Sec.~\ref{sec:geometry}, which reverses $\sigma_z$.
These statements apply to the specified effective representation; they
do not supply a microscopic derivation for molecular orbitals.

\bibliography{bibliography}

\clearpage
\onecolumngrid
\setlength{\tabcolsep}{6pt}
\section*{Supplementary Material}
\makeatletter
\@removefromreset{equation}{section}
\makeatother
\setcounter{section}{0}
\setcounter{equation}{0}
\setcounter{figure}{0}
\setcounter{table}{0}
\renewcommand{\thesection}{S\arabic{section}}
\renewcommand{\theequation}{S\arabic{equation}}
\renewcommand{\thefigure}{S\arabic{figure}}
\renewcommand{\thetable}{S\arabic{table}}
\renewcommand{\theHequation}{supp.\arabic{equation}}
\renewcommand{\theHfigure}{supp.\arabic{figure}}
\renewcommand{\theHtable}{supp.\arabic{table}}
\renewcommand{\theHsection}{supp.\arabic{section}}

\section{Basis and reservoir currents}
The ordered single-particle basis is
$(A1\uparrow,A1\downarrow,B1\uparrow,B1\downarrow,
A2\uparrow,\ldots,BN\downarrow)$.
All energy values below are in meV. Each lead has two orbital modes and two
spin states. Each probe has four channels, one per state of a rung, and a common
spin-independent occupation. For a reservoir with retarded self-energy
$-ig_qP_q$, the level-width matrix is $\Gamma_q=2g_qP_q$.

The Hamiltonian separates as $H_0+H_{\mathrm{so}}$. The real scalar rung
hopping $\gamma_{\mathrm{out}}$ belongs to $H_0$, not to
$H_{\mathrm{so}}$. Appendix B of the main text derives the allowed rung
matrix under TRS and the half-turn at relative azimuth $\beta=\pi$ in
this effective basis. The scalar term is allowed; additional allowed
spin-dependent rung terms are set to zero. This effective restriction is
not a microscopic projection of molecular orbitals.

For distinct terminals define
\begin{equation}
 C_{pq}(E)=\operatorname{Tr}\bigl[\Gamma_pG^r\Gamma_qG^a\bigr],\qquad
 Z_{Rq}(E)=\operatorname{Tr}\bigl[\sigma_z\Gamma_RG^r\Gamma_qG^a\bigr].
\end{equation}
Scattering unitarity gives the charge and outgoing spin currents as
\begin{align}
 I_p^{\mathrm{out}}&=\frac{e}{h}\int dE\sum_{q\ne p}
 C_{pq}(E)[f_p(E)-f_q(E)],\\
 I_{z,R}&=\frac{e}{h}\int dE\sum_{q\ne R}
 Z_{Rq}(E)[f_q(E)-f_R(E)].
\end{align}
The spin expression is equivalent to including the reflection block of the
scattering matrix: its equilibrium contribution cancels the other terms in
the same outgoing row. Omitting reinjection does not implement either type
of zero-charge-current probe.

The reference physical leads have $\mu_L=2.5$ and $\mu_R=-1.5$.
Below $\mu_R$ all occupations equal one, and above $\mu_L$ they vanish.
Those regions make no contribution to the conserving currents.
The internal Hamiltonian is fixed under bias. The numerical calculation does
not solve electrostatics or a microscopic electron--phonon problem.

\section{Elastic cancellation and thermalized occupations}
Let $S|n,j,s\rangle=(-1)^{n+j}|n,j,s\rangle$ and
$Y=\mathbf{1}_{2N}\otimes\sigma_y$. For zero onsite energies,
$SHS=-H$ and $YH^*Y=H$. Spin-independent diagonal reservoir couplings commute
with $S$. Consequently
\begin{equation}
 G^r(-E)=-SG^a(E)S,\qquad G^r(E)=Y[G^r(E)]^TY.
\end{equation}
Restricting $S$ to the terminal channels gives matrices $D_p$. The enlarged
scattering matrix obeys
\begin{equation}
 \mathcal S_{pq}(-E)=D_p\mathcal S_{qp}^{\dagger}(E)D_q
 =D_pY_p\mathcal S_{pq}^*(E)Y_qD_q.
\end{equation}
The charge trace is even, while the outgoing $z$-spin trace is odd because
$Y_p\sigma_zY_p=-\sigma_z$. These identities hold for reflection blocks too.
With $\mathcal M_{pq}=4\delta_{pq}-\mathcal T_{pq}$, the elastic distribution
inside the physical-lead window is
\begin{equation}
 \bm f_P(E)=-\mathcal M_{PP}^{-1}(E)\mathcal M_{PL}(E).
\end{equation}
It is even in energy for the connected, onsite-free model. Thus the effective
elastic spin integrand is odd and a centered window gives exactly zero.
No finite-grid integration is needed to prove that cancellation.

For thermalized probes, the charge constraints are imposed only after
integration, and the occupations are $\theta(\mu_p-E)$ at zero temperature.
They need not be even inside a centered physical-lead window. The calculated
response at $E_F=0$, $\Delta\mu=4$, $g_p=0.05$, and $N=M=10$ is
$I_{z,R}/I_* = \CenteredSpin$ and $P_{\mathrm{int}}=\CenteredPPercent\%$.
The enantiomer reverses the spin signal and has the same charge.
Table~\ref{tab:mu} gives the chemical potentials, showing explicitly that the
probe occupations are not all identical central Fermi steps.

\begin{table}[ht]
\centering
\caption{Calculated thermalized-probe chemical potentials at $g_p=0.05$.
The two columns change only the physical-lead window center; all device
parameters remain fixed.}\label{tab:mu}
\begin{tabular}{rrr}
\toprule
Rung $n$ & $\mu_p$, $E_F=0.5$ & $\mu_p$, $E_F=0$\\
\midrule
1 & 1.025114572 & 0.696260863\\
2 & 0.423108337 & 0.160221553\\
3 & 0.446239084 & 0.200632543\\
4 & 0.400377432 & 0.130280404\\
5 & 0.284267164 & 0.016936429\\
6 & 0.198584752 & -0.016936429\\
7 & 0.152564657 & -0.130280404\\
8 & 0.063848746 & -0.200632543\\
9 & 0.100364023 & -0.160221553\\
10 & -0.360054636 & -0.696260863\\
\bottomrule
\end{tabular}

\par\medskip{\footnotesize\raggedright\noindent\textbf{Alternative text:} The ten probe chemical potentials are listed for displaced and centered physical-lead windows at the same coupling. They are not all equal to the central energy, so their individual Fermi steps do not share the even occupation of elastic probes.\par}
\end{table}

\section{Finite-bias comparison and numerical accuracy}
Table~\ref{tab:environment} reports spin current and normalized polarization
for the same displaced physical-lead window in all prescriptions. An increase
in spin-current magnitude is not the same as an increase in polarization,
because the transmitted charge also changes. In the coherent reference,
$I_0/I_*=\RefCharge$ and $I_z/I_*=\RefSpin$.
The sign change at strong thermalized coupling is a sampled model result,
not a claim of a universal threshold.

\begin{table}[ht]
\centering\small
\caption{Equal-window environmental comparison, $E_F=0.5$,
$\Delta\mu=4$, and $N=M=10$. $g$ denotes $\eta_d$ for loss only and $g_p$
for the probes. The listed spin currents are multiplied by $10^3$.}
\label{tab:environment}
\begin{tabular}{rlrrr}
\toprule
$g$ & Prescription & $I_0/I_*$ & $10^3 I_z/I_*$ & $P_{\rm int}$ (\%)\\
\midrule
0.01 & Loss only & 8.762460 & -4.670748 & -0.053304\\
0.01 & Elastic & 9.449594 & -3.977620 & -0.042093\\
0.01 & Thermalized & 9.474614 & -16.055637 & -0.169460\\
0.025 & Loss only & 7.330731 & -6.026738 & -0.082212\\
0.025 & Elastic & 8.831368 & -4.619556 & -0.052308\\
0.025 & Thermalized & 8.877037 & -27.672338 & -0.311729\\
0.05 & Loss only & 5.492220 & -7.136924 & -0.129946\\
0.05 & Elastic & 7.973570 & -5.197721 & -0.065187\\
0.05 & Thermalized & 8.033499 & -35.587438 & -0.442988\\
0.1 & Loss only & 3.128627 & -6.969854 & -0.222777\\
0.1 & Elastic & 6.684939 & -5.342393 & -0.079917\\
0.1 & Thermalized & 6.756813 & -34.401315 & -0.509135\\
0.25 & Loss only & 0.614930 & -2.916263 & -0.474243\\
0.25 & Elastic & 4.496119 & -3.685718 & -0.081976\\
0.25 & Thermalized & 4.574349 & -17.001697 & -0.371675\\
0.5 & Loss only & 0.046992 & -0.381677 & -0.812222\\
0.5 & Elastic & 2.880294 & -1.564676 & -0.054323\\
0.5 & Thermalized & 2.940658 & -3.807269 & -0.129470\\
1 & Loss only & 0.000430 & -0.005661 & -1.315405\\
1 & Elastic & 1.617195 & -0.318488 & -0.019694\\
1 & Thermalized & 1.649662 & 0.090991 & 0.005516\\
\bottomrule
\end{tabular}

\par\medskip{\footnotesize\raggedright\noindent\textbf{Alternative text:} For each sampled environmental coupling, the table lists charge current, scaled signed spin current, and polarization for loss-only, elastic, and thermalized prescriptions. These observables distinguish signal enhancement from charge normalization and include a sampled sign change.\par}
\end{table}

Coherent and loss-only currents use adaptive vector quadrature with absolute
and relative targets $10^{-9}$. Elastic-probe integration targets $10^{-8}$.
These targets are numerical error estimates, not rigorous interval bounds.
The coherent reference has a quadrature error estimate $5.61\times10^{-10}$
in the integrated two-component current vector. No extra finite numerical
imaginary part is used in the coherent resolvent.

Thermalized potentials are determined using cumulative integrals of cubic
spectral interpolants on successively refined grids of 1001, 2001, 4001,
and, if needed, 8001 points. Refinement stops at 2001 points or more when the
change in both charge and spin current is less than $10^{-9}$.
The resulting potentials are then tested by direct adaptive quadrature,
split at the probe Fermi steps, with targets $2\times10^{-9}$.
The largest integrated probe-current residual in the displaced-window sweep
is \ProbeResidual\ in units $I_*$.

A separate rail-first Hamiltonian and direct Green-function block traces
reproduce the charge and spin currents at $g_p=0.05$ and $1$, and at the
centered operating point, with maximum current difference
$3.56\times10^{-15} I_*$. The corresponding adaptive quadrature estimates
are at most $1.44\times10^{-9} I_*$. The agreement checks implementation
consistency; it does not replace the quadrature error estimates.

\section{Disorder sampling}
Each onsite energy is sampled independently and uniformly from $[-W/2,W/2]$.
There are 128 realizations for each $(N,W)$, with $N=6,10,20$ and
$W=0.01,0.1,0.5,1,2$. Charge and spin currents are integrated for each
realization before any averaging. Each onsite array is retained, so an
individual realization can be recalculated without relying on a random seed.
The recorded seed sequence is $(20261015,N,10^4W)$.

For currents $x_r=I_z^{(r)}/I_*$, the signed sample mean is
$\bar x=\sum_r x_r/128$ and its standard error is $s/\sqrt{128}$, with
sample standard deviation $s$. Main-text error bars are two-sided 95\%
Student-$t$ intervals with 127 degrees of freedom, separately at each point.
They are not simultaneous confidence bands across the sweep.
Percentile bootstrap intervals from 4096 resamples are retained separately.
The mean absolute current, mean polarization, and ratio of mean currents are
distinct statistics and are stored as such.

\section{Conservation residuals and disorder partners}
The elastic-probe conditions were evaluated on 161 uniformly spaced energies
in $[-1.5,2.5]$ meV for every length and coupling used in the main-text
environmental and length sweeps. Define the dimensionless spectral charge
current $j_p(E)=\sum_q\mathcal M_{pq}(E)f_q(E)$. Table~\ref{tab:elasticresidual}
lists the largest sampled probe residual for each length. The probe
occupations remain between zero and one. These sampled residuals test the
implemented constraints; they are not bounds at unsampled energies.

An independent rail-first construction of the Hamiltonian, direct inversion
of its resolvent, and Caroli block traces reproduce the elastic integrands
and integrated currents. This evaluation uses pairwise charge-current
differences and includes every probe contribution to the outgoing spin
current. It does not construct the enlarged scattering matrix.
The last column of Table~\ref{tab:elasticresidual} compares its integrals
with the retained currents. Both charge and spin are included in the maximum.

\begin{table}[ht]
\centering\small
\caption{Elastic-probe charge residuals and independent-current comparison.
The discrete set $\mathcal G=\{0.01,0.025,0.05,0.1,0.25,0.5,1\}$ is in meV;
the other rows use $g_p=0.05,0.1$ meV. The energy window and all other
parameters are those of the reference device.}
\label{tab:elasticresidual}
\begin{tabular}{rccc}
\toprule
$N$ & $g_p$ (meV) & $\max |j_p(E)|$ & $\max |\Delta I_{0,z}|/I_*$ \\
\midrule
6 & $0.05,0.1$ & $2.78\times10^{-16}$ & $5.33\times10^{-15}$ \\
8 & $0.05,0.1$ & $2.81\times10^{-16}$ & $3.55\times10^{-15}$ \\
10 & $\mathcal G$ & $1.11\times10^{-15}$ & $1.60\times10^{-14}$ \\
14 & $0.05,0.1$ & $3.33\times10^{-16}$ & $7.11\times10^{-15}$ \\
20 & $0.05,0.1$ & $5.10\times10^{-16}$ & $7.11\times10^{-15}$ \\
26 & $0.05,0.1$ & $4.44\times10^{-16}$ & $9.77\times10^{-15}$ \\
30 & $0.05,0.1$ & $5.55\times10^{-16}$ & $1.33\times10^{-14}$ \\
\bottomrule
\end{tabular}
\par\medskip{\footnotesize\raggedright\noindent\textbf{Alternative text:} Elastic-probe constraints are tested across the published lengths and couplings on the stated energy mesh. Maximum spectral charge residuals, occupation bounds, and independent integrated-current differences quantify the sampled consistency checks.\par}
\end{table}

For voltage probes, Table~\ref{tab:voltageresidual} gives the integrated
current into each probe, maximized over the probes, using the retained
chemical potentials and independent adaptive Caroli quadrature. The
quadrature is split at every probe Fermi step and uses absolute and relative
targets $2\times10^{-10}$. These are numerical targets, not rigorous
enclosures. The potentials for all listed cases stabilized at 2001 spectral
points under the previously stated refinement rule. Their charge constraints
are integrated constraints, not energy-resolved nulls.

\begin{table}[ht]
\centering\small
\caption{Voltage-probe residuals and independent-current comparison.
$E_F$ and $g_p$ are in meV; $\Delta\mu=4$ meV and $N=M=10$ throughout.
The last column compares both physical charge and spin currents with the
retained values. Full per-probe current vectors accompany the numerical data.}
\label{tab:voltageresidual}
\begin{tabular}{cccc}
\toprule
$E_F$ & $g_p$ & $\max_p |I_p|/I_*$ & $\max |\Delta I_{0,z}|/I_*$ \\
\midrule
0.5 & 0.01 & $5.50\times10^{-12}$ & $3.55\times10^{-15}$ \\
0.5 & 0.025 & $1.99\times10^{-11}$ & $1.94\times10^{-16}$ \\
0.5 & 0.05 & $3.40\times10^{-12}$ & $5.33\times10^{-15}$ \\
0.5 & 0.1 & $8.10\times10^{-12}$ & $3.55\times10^{-15}$ \\
0.5 & 0.25 & $2.45\times10^{-12}$ & $1.78\times10^{-15}$ \\
0.5 & 0.5 & $4.27\times10^{-13}$ & $8.88\times10^{-16}$ \\
0.5 & 1 & $6.69\times10^{-14}$ & $8.88\times10^{-16}$ \\
0 & 0.05 & $2.20\times10^{-12}$ & $1.78\times10^{-15}$ \\
0 & 0.1 & $2.78\times10^{-12}$ & $2.66\times10^{-15}$ \\
\bottomrule
\end{tabular}
\par\medskip{\footnotesize\raggedright\noindent\textbf{Alternative text:} Voltage-probe constraints are tested in displaced and centered windows. The table reports maximum integrated probe-current residuals and differences in physical currents from an independent Green-function evaluation; these are integrated rather than energy-resolved constraints.\par}
\end{table}

For onsite disorder, the partner operation is global sign inversion of the
entire onsite pattern, $\epsilon\mapsto-\epsilon$, together with reversal
of energy or of the window center. Appendix A of the main text derives
$I_z(-E_F;-\epsilon)=-I_z(E_F;\epsilon)$ for the stated coherent model.
The relation does not imply a cancellation of the pair at the same displaced
window. For a centered window the two partner currents cancel. Direct
integration of one retained pattern from each $(N,W)$ group and its sign
partner gives a largest centered-pair residual of $1.68\times10^{-15}$ in units $I_*$.
These constructed partners are symmetry controls, not extra independent
realizations for the reported confidence intervals.

The signed mean, its Student-$t$ interval, and the mean absolute current
answer different questions. For example, at $N=10$, $W=1$ meV and the
displaced reference window, the mean absolute current is $0.0367173$ $I_*$,
whereas the signed mean is $-0.0013238$ $I_*$ with 95\% interval
$[-0.0096592,0.0070116]I_*$. The sample mean absolute current is finite, whereas
this interval does not resolve a nonzero signed population mean. It does not
establish an exactly zero population mean either. The interpretation remains
separate from the exact ensemble cancellation at $E_F=0$.

\clearpage
\section{Current-based control plots}
\begin{figure}[!ht]
\centering
\includegraphics[width=\textwidth]{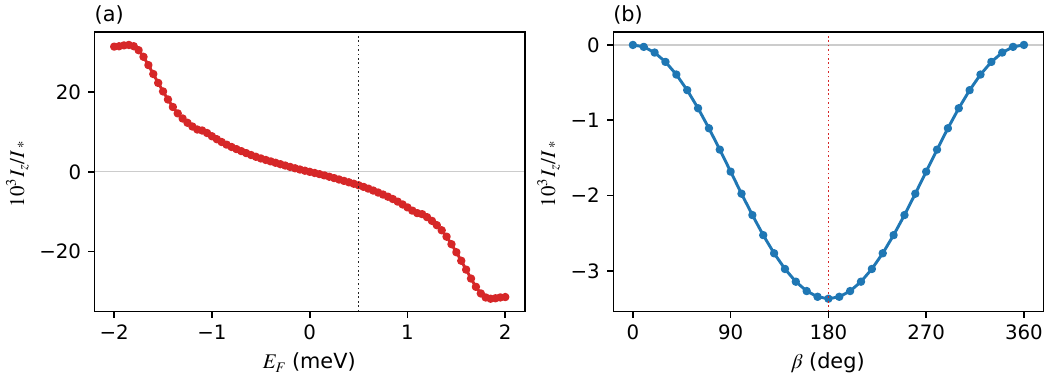}
\caption{Integrated spin current versus (a) window center at
$\Delta\mu=4$ meV and (b) strand offset at $E_F=0.5$ meV, for $N=M=10$.
Dotted vertical lines mark $E_F=0.5$ meV and $\beta=180^\circ$.
All remaining parameters are those of the reference device.}
\label{fig:currentcontrols}
\par\medskip{\footnotesize\raggedright\noindent\textbf{Alternative text:} Current-based controls show how the signed spin signal changes with transport-window center and relative strand angle. The window offset reverses the signal, and equivalent strand angles give zero, before division by the transmitted charge.\par}
\end{figure}

Figure~\ref{fig:currentcontrols} shows the currents underlying the main-text
polarization sweeps. The offset reverses the sign, and equivalent-strand
angles give nulls. Polarization additionally requires charge normalization.

\begin{figure}[!ht]
\centering
\includegraphics[width=\textwidth]{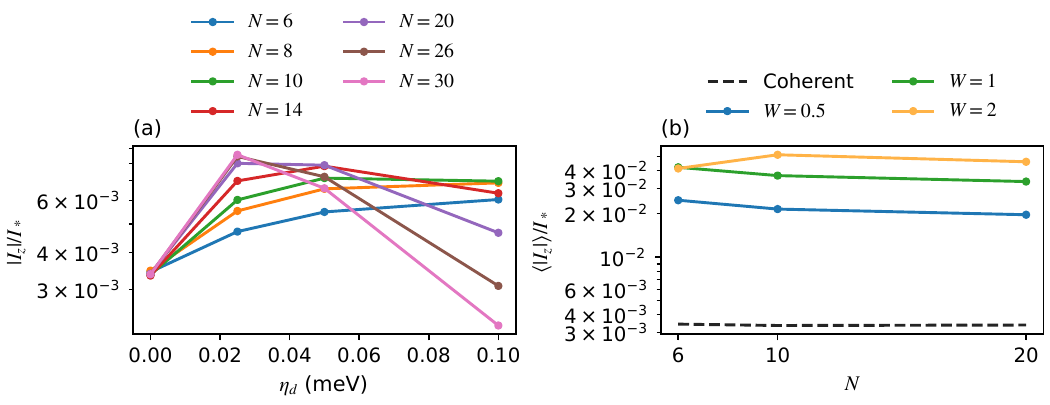}
\caption{(a) Loss-only spin-current magnitude versus absorber halfwidth,
with sampled $\eta_d=0,0.025,0.05,0.1$ meV at each indicated length.
(b) Mean absolute spin current versus length at $W=0.5,1,2$ meV and the
coherent reference. The disorder means use 128 realizations per point.
Both panels use $M=10$, $E_F=0.5$ meV and $\Delta\mu=4$ meV. Lines connect
sampled points; the scan in (a) does not locate a global coupling maximum.}
\label{fig:environmentcontrols}
\par\medskip{\footnotesize\raggedright\noindent\textbf{Alternative text:} Loss-only spin-current magnitudes are plotted against absorber coupling for the sampled ladder lengths. A second panel compares coherent current magnitudes and disorder mean absolute currents versus length, without asserting a resolved nonzero signed disorder mean.\par}
\end{figure}

Figure~\ref{fig:environmentcontrols} separates loss-only absorption from
the typical magnitude of a disordered coherent sample. The absolute means
in (b) cannot determine the sign or establish a zero signed population mean;
the main-text signed means and uncertainty supply that distinct information.

\end{document}